\documentclass[reprint,superscriptaddress,amsmath,amssymb,aps,prx,longbibliography]{revtex4-1}

\usepackage{graphicx}
\usepackage{dcolumn}
\usepackage{bm}
\usepackage[colorlinks=true,urlcolor=blue,linkcolor=blue,citecolor=blue]{hyperref}
\usepackage[normalem]{ulem}
\usepackage{tikz}
\usetikzlibrary{quantikz}
\usepackage{amsmath}
\usepackage{amssymb}
\usepackage{bbold}

\usepackage{enumitem}
\usepackage{algpseudocode}
\usepackage{algorithm}
\usepackage{natbib}
\DeclareUnicodeCharacter{03B2}{\ensuremath{\beta}}
\DeclareUnicodeCharacter{2009}{\,}
 
\usepackage[all]{hypcap} 
\usepackage{braket}

\newcommand{\customref}[2]{\hyperref[#1]{\ref*{#1}#2}}

\usepackage{xcolor}

\definecolor{Ured}{HTML}{FF5C5C}
\definecolor{Ublue}{HTML}{ADD8E6}
\definecolor{Ugreen}{HTML}{198a11}

\newcommand{\xuparrow}[1]{%
  {\left\uparrow\vbox to #1{}\right.\kern-\nulldelimiterspace}
}

\begin{document}

\title{Unitary fault-tolerant encoding of Pauli states in surface codes}

\author{Luis Colmenarez}
\email{colmenarez@physik.rwth-aachen.de}
\affiliation{Institute for Theoretical Nanoelectronics (PGI-2), Forschungszentrum Jülich, 52428 Jülich, Germany}
\affiliation{Institute for Quantum Information, RWTH Aachen University, 52056 Aachen, Germany}

\author{Remmy Zen}
\affiliation{Max Planck Institute for the Science of Light, Staudtstra{\ss}e 2, 91058 Erlangen, Germany}
\affiliation{School of Physics and Astronomy, Monash University, Clayton, VIC 3168, Australia}

\author{Jan Olle}
\affiliation{Max Planck Institute for the Science of Light, Staudtstra{\ss}e 2, 91058 Erlangen, Germany}
\affiliation{NVIDIA Corporation, 2788 San Tomas Expressway, Santa Clara, 95051, CA, USA}

\author{Florian Marquardt}%
\affiliation{Max Planck Institute for the Science of Light, Staudtstra{\ss}e 2, 91058 Erlangen, Germany}
\affiliation{Department of Physics, Friedrich-Alexander Universit\"{a}t Erlangen-N\"{u}rnberg, Staudtstra{\ss}e 5, 91058 Erlangen, Germany}

\author{Markus M\"{u}ller}%
\affiliation{Institute for Theoretical Nanoelectronics (PGI-2), Forschungszentrum Jülich, 52428 Jülich, Germany}
\affiliation{Institute for Quantum Information, RWTH Aachen University, 52056 Aachen, Germany}

\date{\today}

\begin{abstract}

In fault-tolerant quantum computation, the preparation of logical states is a ubiquitous subroutine, yet significant challenges persist even for the simplest states required.
In the present work, we present a unitary, scalable, distance-preserving encoding scheme for preparing Pauli eigenstates in surface codes. Unlike previous unitary approaches whose fault-distance remains constant with increasing code distance, our scheme ensures that the protection offered by the code is preserved during state preparation. 
Building on strategies discovered by reinforcement learning for the surface-17 code, we generalize the construction to arbitrary code distances and both rotated and unrotated surface codes. 
The proposed encoding relies only on geometrically local gates, and is therefore fully compatible with planar 2D qubit connectivity, and it achieves circuit depth scaling as $O(d)$, consistent with fundamental entanglement-generation bounds. We design explicit stabilizer-expanding circuits with and without ancilla-mediated connectivity and analyze their error-propagation behavior. Numerical simulations under depolarizing noise show that our unitary encoding without ancillas outperforms standard stabilizer-measurement-based schemes, reducing logical error rates by up to an order of magnitude. These results make the scheme particularly relevant for platforms such as trapped ions and neutral atoms, where measurements are costly relative to gates and idling noise is considerably weaker than gate noise. Our work bridges the gap between measurement-based and unitary encodings of surface-code states and opens new directions for distance-preserving state preparation in fault-tolerant quantum computation.

\end{abstract}
\maketitle

\section{Introduction}

Quantum error correction (QEC) promises to enable fault-tolerant (FT) quantum computation by correcting errors faster than they accumulate. QEC achieves this by introducing redundancy through the use of additional qubits, which allows for the detection and correction of errors. Consequently, many physical qubits are used to encode one or more logical qubits, which are the carriers of quantum information in FT quantum algorithms.
Recent experiments across several platforms—including superconducting qubits~\cite{besedin_realizing_2025, huang_logical_2025,google_quantum_ai_and_collaborators_quantum_2025}, cat qubits~\cite{putterman_hardware-efficient_2025}, trapped ions~\cite{pogorelov_experimental_2025, yamamoto_quantum_2025}, and neutral atoms~\cite{sales_rodriguez_experimental_2025, muniz_repeated_2025, bluvstein_architectural_2025}—have demonstrated the successful implementation of key QEC protocols, along with the expected suppression of errors in the low-noise regime~\cite{google_quantum_ai_and_collaborators_quantum_2025, bluvstein_logical_2024}.
Today, a leading contender for practical FT QEC are topological codes \cite{kitaev_fault-tolerant_2003}. Among them, the surface code~\cite{kitaev_fault-tolerant_2003, dennis_topological_2002, fowler_surface_2012, fowler_high-threshold_2009} is the most widely studied and experimentally pursued one, owing to its high threshold under circuit-level noise, its compatibility with a nearest neighbor layout in 2D, and its well-understood framework for scalable fault-tolerance.
Several critical components for FT quantum computation are already well established within surface code architectures. Decoding can be performed efficiently using the minimum weight perfect matching (MWPM) algorithm~\cite{dennis_topological_2002, higgott_pymatching_2021}, while maintaining a relatively high error threshold~\cite{fowler_high-threshold_2009}. Logical CNOT gates can be implemented transversally, a strategy already demonstrated in neutral atom platforms~\cite{bluvstein_logical_2024}, or via lattice surgery~\cite{horsman_surface_2012}, a method especially relevant when qubit interactions are restricted to nearest neighbors, as is the case in superconducting qubit architectures~\cite{krinner_realizing_2022, zhao_realization_2022, google_quantum_ai_and_collaborators_quantum_2025}.
Single-qubit logical gates can be executed through fold-transversal gate constructions~\cite{breuckmann_fold-transversal_2024, moussa_transversal_2016, kubica_unfolding_2015}, or by employing magic state distillation and injection~\cite{sales_rodriguez_experimental_2025, postler_demonstration_2022, bravyi_universal_2005}. More recently, magic state cultivation has emerged as a promising alternative~\cite{gidney_magic_2024,sahay_fold-transversal_2025}.

One fundamental operation on logical qubits is state preparation and encoding~\cite{gottesman_stabilizer_1997}. This task involves preparing a desired logical state through the application of gates and measurements on bare physical qubits. As any protocol involving logical qubits begins with the preparation of one or more logical states, this operation is ubiquitous across all QEC schemes and quantum algorithms.
However, preparing logical states is particularly challenging due to the complex many-body entanglement structure inherent to these states, especially when device-specific constraints are taken into account. For example, in topological QEC codes, encoding logical states is equivalent to preparing topologically ordered states~\cite{liao_graph-state_2021, aguado_entanglement_2008, aharonov_quantum_2018}. Consequently, any unitary circuit that prepares logical states in surface codes must respect limits on how quickly entanglement can be generated, as dictated by Lieb–Robinson-type bounds~\cite{bravyi_lieb-robinson_2006}.
Specifically, when only geometrically local gates are allowed, the circuit depth for encoding scales as $\mathcal{O}(d)$, where $d$ is both the code distance and the linear size of the lattice~\cite{higgott_optimal_2021, chen_quantum_2024}. In contrast, if all-to-all connectivity is available, the depth can be reduced to $\mathcal{O}(\log d)$~\cite{aguado_entanglement_2008, vidal_class_2008, liao_graph-state_2021, tsai_unitary_2025}. 
In contrast, measurement-based schemes can generate entanglement more rapidly, enabling logical state preparation in constant time~\cite{dennis_topological_2002,raussendorf_long-range_2005,bergamaschi_fault_2025}.

Another important factor to consider is the resilience of encoding schemes to errors occurring during gate execution and measurements. Although these errors are initially local, they can propagate rapidly as more entangling gates are applied~\cite{dennis_topological_2002}. This vulnerability is especially pronounced in previously proposed unitary encoding circuits for surface codes~\cite{aguado_entanglement_2008, vidal_class_2008, liao_graph-state_2021, tsai_unitary_2025}. In such schemes, the \emph{fault-distance}, defined as the minimum number of faults required to produce an undetectable logical error, remains constant regardless of the \emph{code distance} $d$. This limitation arises from the uncontrolled spread of errors during circuit execution.
To date, the only known \emph{distance-preserving} scalable encoding scheme is the one in which qubits are initialized on a product state, e.g. $|0\rangle$, and stabilizers of the opposite type, e.g. $X-$type, are measured at least once~\cite{bluvstein_logical_2024}, under the assumption that the stabilizer measurement circuits themselves are FT. This method ensures that the fault-distance is equal to the code distance, preserving during state preparation the protection offered by the QEC code.
Some encoding methods rely on state verification using flag qubits \cite{chao_flag_2020,chamberland_flag_2018,chao_quantum_2018}. While this can make the encoding distance-preserving, such schemes are difficult to scale beyond small distances.
It is important to note that when measurements and classical feed-forward are allowed the circuit depth to prepare a fully FT state can be constant \cite{dennis_topological_2002,bergamaschi_fault_2025,raussendorf_long-range_2005,cobos_noise-aware_2024}. In this work we focus on encoding schemes that do not rely on classical feed-forward, e.g. decoding, to achieve fault-tolerance or distance preservation.

In this work, we propose, for the first time, a unitary scalable encoding for the surface code that is distance-preserving, inspired by our recent results in Ref.~\cite{zen_quantum_2025}, where a reinforcement learning agent was used to find a FT encoding circuit for the surface-17 code \cite{krinner_realizing_2022,google_quantum_ai_exponential_2021,tomita_low-distance_2014}. Interestingly, the agent discovered a novel strategy that was previously unknown to human experts. This further illustrates how AI can provide new insights for solving complex quantum tasks. Here, we generalize the strategy that the reinforcement learning agent discovered to arbitrary code distances and to the preparation of Pauli eigenstates $|0\rangle_L$ and $|+\rangle_L$ for both rotated \cite{tomita_low-distance_2014} and unrotated surface codes \cite{dennis_topological_2002}.
We assume a standard square-grid architecture of data and ancilla qubits, typical of superconducting qubit arrays~\cite{google_quantum_ai_and_collaborators_quantum_2025, krinner_realizing_2022}, where each stabilizer is associated with an ancilla placed at its center. We design unitary encoding circuits where two-qubit gates are applied either between data and ancilla qubits (see Fig.~\ref{fig:surface_code}a and \ref{fig:surface_code}b) or, in a variant of this scheme, between data qubits within the same plaquette (see Fig.~\ref{fig:surface_code}c and \ref{fig:surface_code}d). The latter is feasible in  architectures with non-static, i.e. dynamically reconfigurable qubit connectivity, such as neutral atoms \cite{bluvstein_logical_2024} and trapped ions \cite{moses_race-track_2023}.
We have tested our circuits on both rotated and unrotated surface codes in the presence of gate errors and find very good performance when compared to logical state preparation based on stabilizer measurement \cite{dennis_topological_2002,krinner_realizing_2022,zen_quantum_2025,google_quantum_ai_and_collaborators_quantum_2025}. 
Importantly, the proposed encoding scheme remains valid in any implementation of the surface code where hook errors can be oriented orthogonally to the desired logical operator and thus remain correctable \cite{tomita_low-distance_2014,old_fault-tolerant_2025,jandura_surface_2024}. 
Furthermore, for QEC codes in which certain classes of high-weight errors remain correctable, our encoding scheme can be adapted to engineer hook errors so that they fall within these correctable error classes.
As expected for unitary encoding schemes constrained to local interactions, the circuit depth scales as $\mathcal{O}(d)$, and is therefore deeper than measurement-based encoding circuits, which can achieve constant depth $\mathcal{O}(1)$.
However, because measurements are typically much slower than gates, our unitary encoding scheme may actually be \emph{faster} than measurement-based encoding, thereby reducing the overall idling time.
For example, in superconducting qubits \cite{google_quantum_ai_and_collaborators_quantum_2025,krinner_realizing_2022}, measurements can take up to five times longer than gates. For neutral atoms the difference can be up to two orders of magnitude \cite{radnaev_universal_2025,henriet_quantum_2020}. Thus, reducing measurements in favor of gate-based operations can significantly shorten the overall QEC protocol and reduce exposure to decoherence.
Furthermore, our scheme is especially relevant for quantum platforms with long coherence times, such as certain trapped ions~\cite{blinov_quantum_2004, ringbauer_universal_2022} and neutral atom arrays~\cite{henriet_quantum_2020}, where idling errors are negligible and measurements are preferably deferred to later stages of computation. 
Thus, implementing our encoding scheme offers a promising path to producing high-quality Pauli states in near-term quantum processors.
Furthermore, our unitary encoding scheme is fully compatible with the paradigm of algorithmic fault-tolerance~\cite{zhou_algorithmic_2024}, making it a promising building block for FT quantum computing architectures that minimize measurement overhead.

The structure of the paper is as follows. In Sec.~\ref{sec:background}, we introduce the basic concepts used in our work. In Sec.~\ref{sec:encoding_FT}, we describe in detail the unitary encoding we develop. In Sec.~\ref{sec:simulations}, we benchmark our unitary encoding scheme against the standard measurement-based encoding. Finally, in Sec.~\ref{sec:conclusions}, we present our concluding remarks.

\section{Background}\label{sec:background}

\subsection{Surface code}

\begin{figure}
    \includegraphics[width=1.0\linewidth]
    {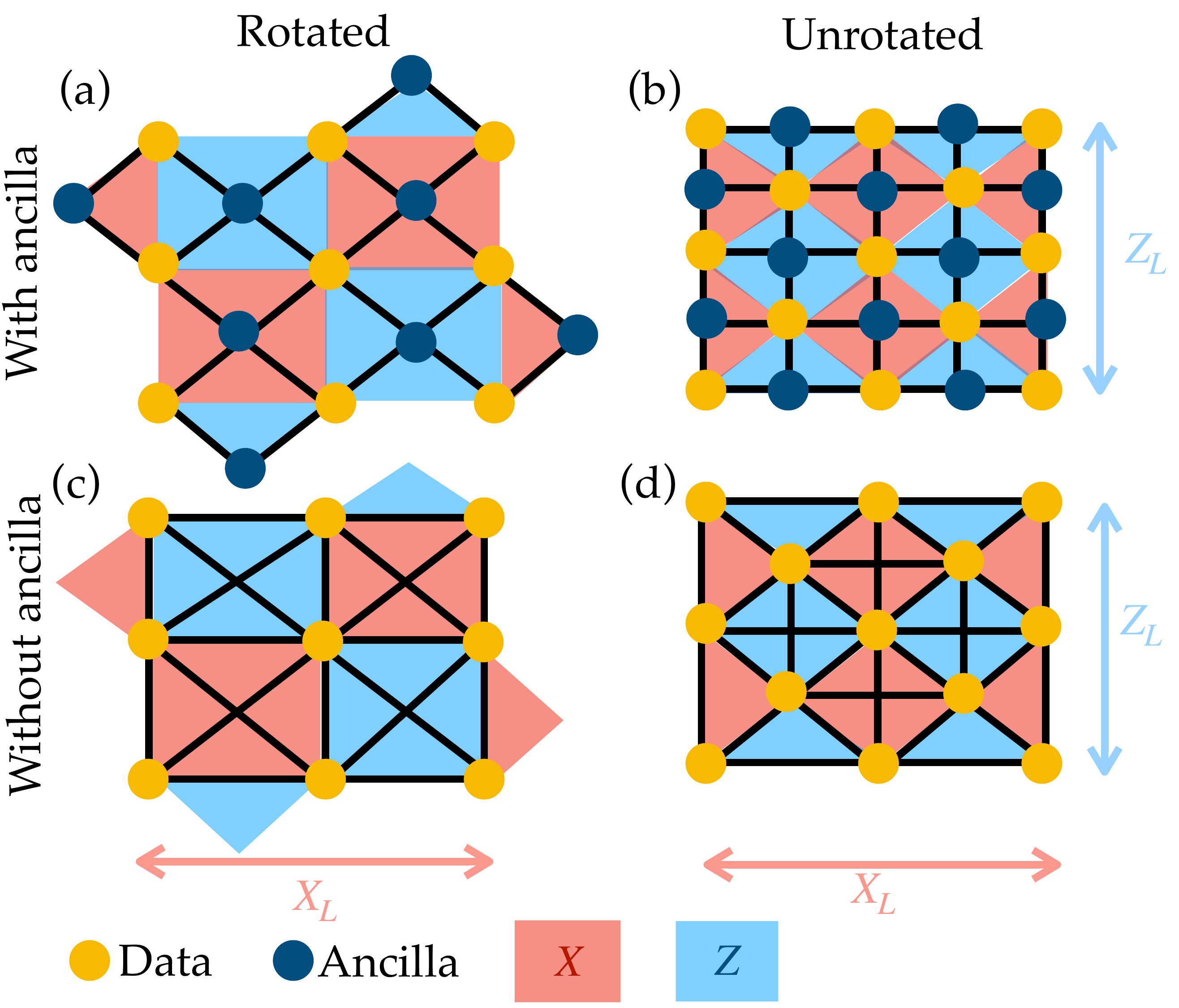}
    \caption{a) and c) Rotated surface code of code distance $d=3$. b) and d) Unrotated surface code of distance $d=3$. Red and blue denote $X$ and $Z$ stabilizers respectively. Logical operators $Z_L$ and $X_L$ are defined along the vertical and horizontal boundary, respectively. Yellow and dark blue dots represent data and ancilla qubits respectively. The black lines represent the qubit connectivity.}
    \label{fig:surface_code}
\end{figure}

The surface code is a planar topological code usually defined on a square lattice~\cite{dennis_topological_2002, tomita_low-distance_2014}. It is characterized by a set of $X$- and $Z$-type stabilizer checks, also referred to as stars and plaquettes, respectively. For simplicity, we assume throughout that all stabilizer operators have eigenvalue $+1$, although this condition can be relaxed as long as the measured eigenvalue is known; see the discussion on measurement-based encoding in Sec.~\ref{subsec:encoding}.
The logical operators $X_L$ and $Z_L$ correspond to string operators that traverse the lattice in the horizontal and vertical directions, respectively (see Fig.~\ref{fig:surface_code}). In this work, we focus on the preparation of logical states that are eigenstates of the corresponding logical operators, satisfying $Z_L |0\rangle_L = |0\rangle_L$ and $X_L |+\rangle_L = |+\rangle_L$.
Importantly, any implementation of the surface code supports efficient decoding using MWPM ~\cite{dennis_topological_2002, higgott_pymatching_2021}.

The most widely used realization of the surface code is the $[[d^2,1,d]]$ \emph{rotated} surface code~\cite{tomita_low-distance_2014, google_quantum_ai_and_collaborators_quantum_2025, krinner_realizing_2022, zhao_realization_2022}, where $d$ denotes both the code distance and the linear size of the lattice. Its main advantage over other realizations is its efficient use of physical qubits, requiring fewer qubits to achieve a given code distance.
The original formulation of the planar surface code, however, is the \emph{unrotated} surface code~\cite{dennis_topological_2002}, which has code parameters $[[d^2 +(d-1)^2, 1, d]]$. This version uses roughly twice as many physical qubits as the rotated code to achieve the same code distance. Nonetheless, recent studies have shown that the unrotated surface code can be more resilient to hook errors in syndrome measurement circuits~\cite{dennis_topological_2002, manes_distance-preserving_2025} when implementing syndrome readout circuits via three-qubit gates~\cite{old_fault-tolerant_2025,jandura_surface_2024}.
These findings motivate a renewed interest in the unrotated surface code, especially in the context of studying stabilizer-related circuits. In Fig.~\ref{fig:surface_code}, we illustrate both the $d=3$ rotated and unrotated surface codes for comparison.

\subsection{Encoding of Pauli states}\label{subsec:encoding}

As a topological code defined on a $d$-dimensional lattice, a surface (or toric) code state can be prepared by a geometrically local unitary circuit of depth $\mathcal{O}(d)$~\cite{aharonov_quantum_2018, higgott_optimal_2021, chen_quantum_2024}. If all-to-all connectivity is available, the circuit depth can be reduced to $\mathcal{O}(\log d)$~\cite{aguado_entanglement_2008, vidal_class_2008, liao_graph-state_2021, tsai_unitary_2025}. Several constructions of unitary encoding circuits for surface codes exist. One general approach exploits the equivalence between graph states and stabilizer states~\cite{liao_graph-state_2021}, enabling a method in which each stabilizer is prepared as a GHZ state. This recipe applies to arbitrary stabilizer codes and does not rely on local connectivity~\cite{amaro2019}.
Other constructions are based on renormalization group techniques, where the typical strategy is to grow a surface code patch from a smaller distance patch~\cite{higgott_optimal_2021, tsai_unitary_2025}. However, all of the aforementioned schemes are limited in their treatment of fault-tolerance and error proliferation. In particular, the encoding schemes proposed in Refs.~\cite{higgott_optimal_2021, tsai_unitary_2025, liao_graph-state_2021, aguado_entanglement_2008} exhibit fault-distances that are lower than the code distance. As a result, the surface code states prepared by these circuits can introduce high-weight logical errors when used as components in logical quantum algorithms. 
To the best of our knowledge, our work is the first to demonstrate a distance-preserving unitary and deterministic encoding scheme for surface codes. 
This scheme assumes only local gate connectivity and achieves circuit depth $\mathcal{O}(d)$. Furthermore, it relies solely on the ability to orient hook errors orthogonal to the desired logical operators, making it applicable to any variant of the surface code. We design explicit encoding circuits that consider gate connectivity within each stabilizer plaquette, both with and without the use of ancilla qubits assigned to measure stabilizers.

It must be noted that our encoding scheme preserves the fault-distance only for the type of error that flips the corresponding codeword, for example, $X$ errors when preparing the logical state $|0\rangle_L$. 
Therefore, a full round of QEC—namely, a fault-tolerant syndrome extraction using repeated stabilizer measurements (or Steane-type syndrome extraction \cite{postler_demonstration_2024}), followed by correction, is required to remove potential harmful errors of the complementary type. Further details are provided in Appendix~\ref{appendix:ft_both}.
This places our unitary encoding circuit on equal footing with the standard measurement-based encoding scheme, in which $|0\rangle_L$ ($|+\rangle_L$) is prepared by initializing all data qubits in the product state $|0\rangle^{\otimes n}$ ($|+\rangle^{\otimes n}$), followed by a single round of $X$- ($Z$-) stabilizer measurements. After this procedure, the $Z$ ($X$) stabilizers are deterministically projected to the $+1$ eigenstate, allowing $X$ ($Z$) errors to be corrected via a round of QEC; that is, by executing FT syndrome extractions and applying the appropriate correction.
However, because the outcomes of the $X$ ($Z$) stabilizer measurements are both random and potentially faulty, a single round of stabilizer measurements does not provide sufficient information to accurately determine the syndrome. This leads to a high chance of introducing a logical $Z_L$ ($X_L$) error, even after an ideal QEC round.  
Fortunately this is not a problem since we are preparing an eigenstate of precisely that logical operator, namely $Z_L|0\rangle_L = |0\rangle_L$ ($X_L|+\rangle_L = |+\rangle_L$). 
Therefore, we choose to compare our unitary encoding scheme directly against the measurement-based one, given their similar properties regarding fault-tolerance.

In summary, achieving a distance-preserving encoding for both types of errors, $X$ and $Z$, requires the execution of a full QEC round following the preparation of the logical state by any of the methods described in this work. Prior to the QEC round, the logical state is FT only with respect to a single type of error, either $X$ or $Z$. 
Therefore, applying FT gadgets, e.g transversal logical CNOTs, \emph{before} this respective QEC round, may end up spreading high-weight errors to other logical qubits. A way to deal with such high-weight errors is to correct multiple logical qubits jointly \cite{zhou_algorithmic_2024,cain_correlated_2024}. However, in this work we consider the standard scenario of correcting each logical qubit independently.

\section{Unitary fault-tolerant encoding}\label{sec:encoding_FT}

\begin{figure*}
    \centering
     \includegraphics[width=0.85\linewidth]
     {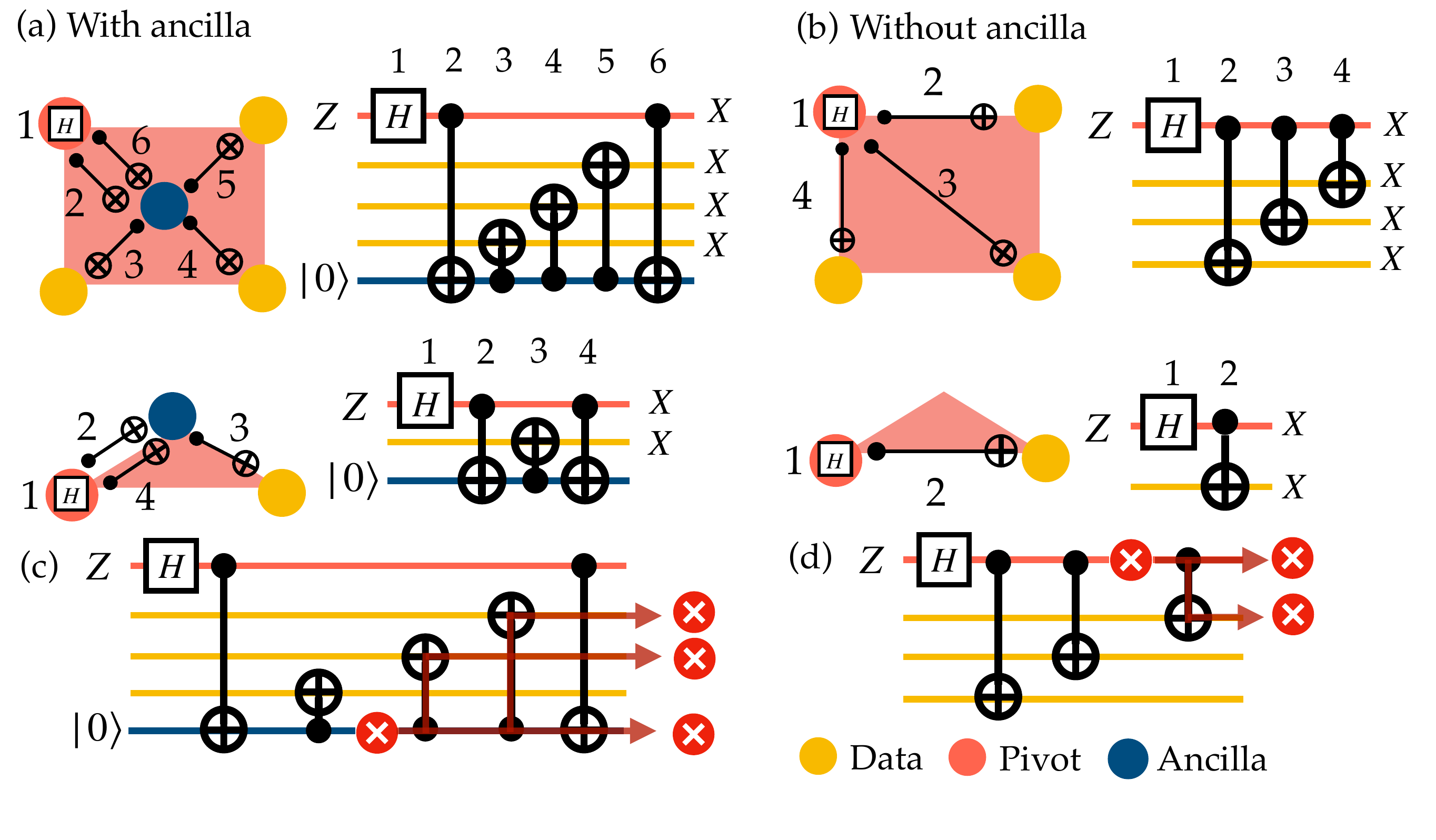}
    \caption{Circuits for expanding the $X$ stabilizers. a) and b) Circuit for coherently initializing an $X$ plaquette with and without ancilla in the middle, respectively. Each circuit assumes that the pivot qubit (positioned at the top of each diagram) is initialized in the state $|+\rangle$ (shown as a Hadamard gate in this depiction). The circuit then transforms the pivot qubit into an $X$-type stabilizer. This procedure is general and applies to stabilizers of any weight, although only weight-two and weight-four examples are shown. Notably, the construction requires at most one ancilla qubit—if any—independent of the stabilizer weight. Numbers indicate the order of the gates to avoid proliferation of dangerous hook errors. c) and d) show the potential hook errors for the circuit realization with and without ancilla, respectively. In addition, several potentially dangerous hook $Z$ errors are discussed in detail in Appendix~\ref{appendix:ft_both}.}
    \label{fig:stab_circuits}
\end{figure*}

\begin{figure*}
    \centering
     \includegraphics[width=1.0\linewidth]{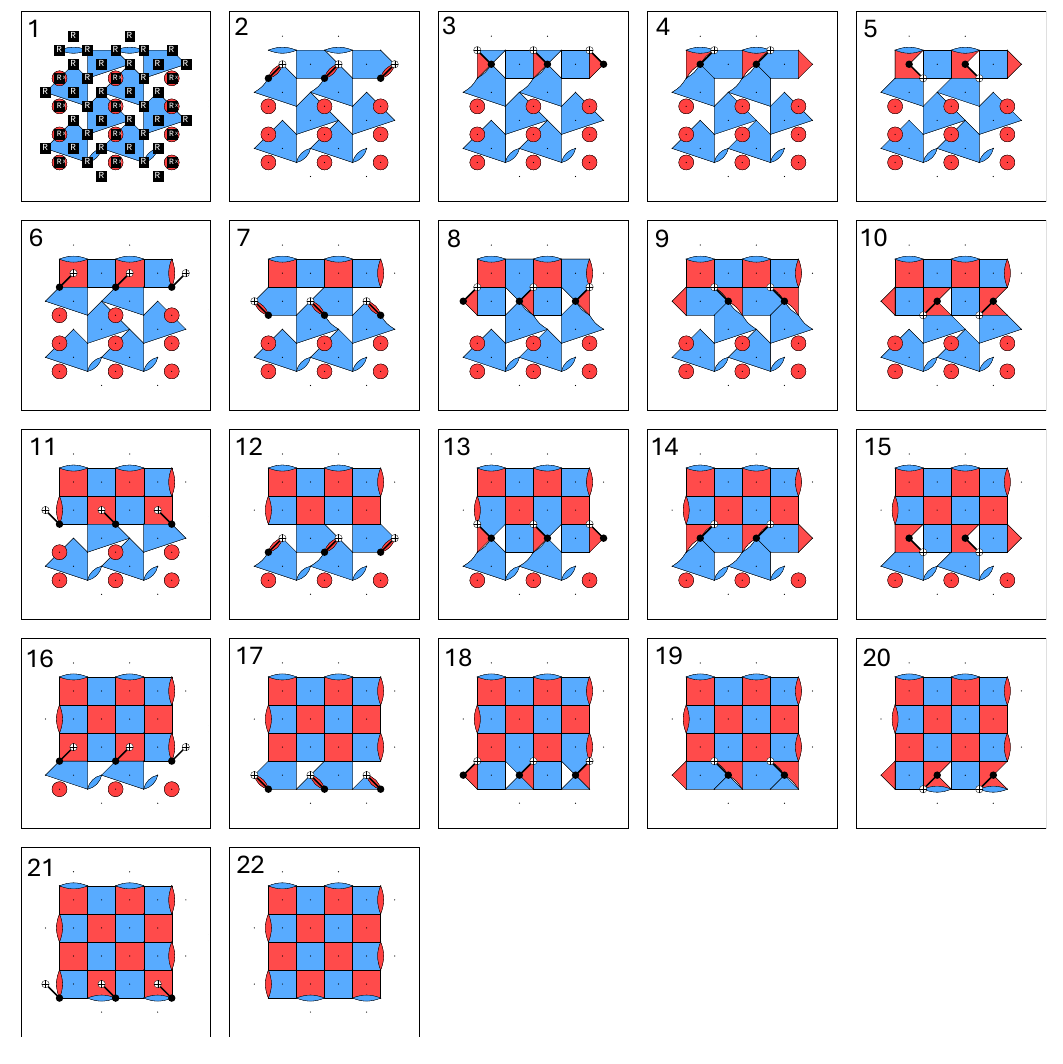}
    \caption{Unitary FT encoding for the $d=5$ rotated surface code using ancillas as bridge qubits. In this layout, blue (red) denotes the $Z$ ($X$) stabilizers and the logical operators $X_L$ and $Z_L$ lie along the horizontal and vertical directions, respectively. Step 1 initializes the data and ancilla qubits in the states $|0\rangle$ except the pivot qubits that are initialized in the state $|+\rangle$.
    The blue irregular regions at this stage denote possible groupings of $|0\rangle$ qubits that will form the standard $Z$ stabilizers at the end of the encoding procedure.
    Steps 2 through 6 prepare the $X$-type stabilizers on the topmost row. Each plaquette within the row can be prepared in parallel. Steps 7 through 11 perform the preparation of $X$ stabilizers on the second row. Steps 12 through 16 prepare the $X$ stabilizers on the third row. Here, the location of the weight-two $X$ stabilizer alternates between the left and right boundaries, such that the gate scheduling appears the same only every second row. Finally, steps 17 through 21 complete the preparation of the $X$ stabilizers on the bottom row. Step 22 depicts the whole set of stabilizers after the last gate is applied. At the end of the circuit, only correctable weight-two $X$ errors remain. However, certain hook $Z$ errors may still be potentially harmful; these are discussed in detail in Appendix~\ref{appendix:ft_both}.}
    \label{fig:circuit_ancilla}
\end{figure*}

\begin{figure*}
    \centering
     \includegraphics[width=1.0\linewidth]{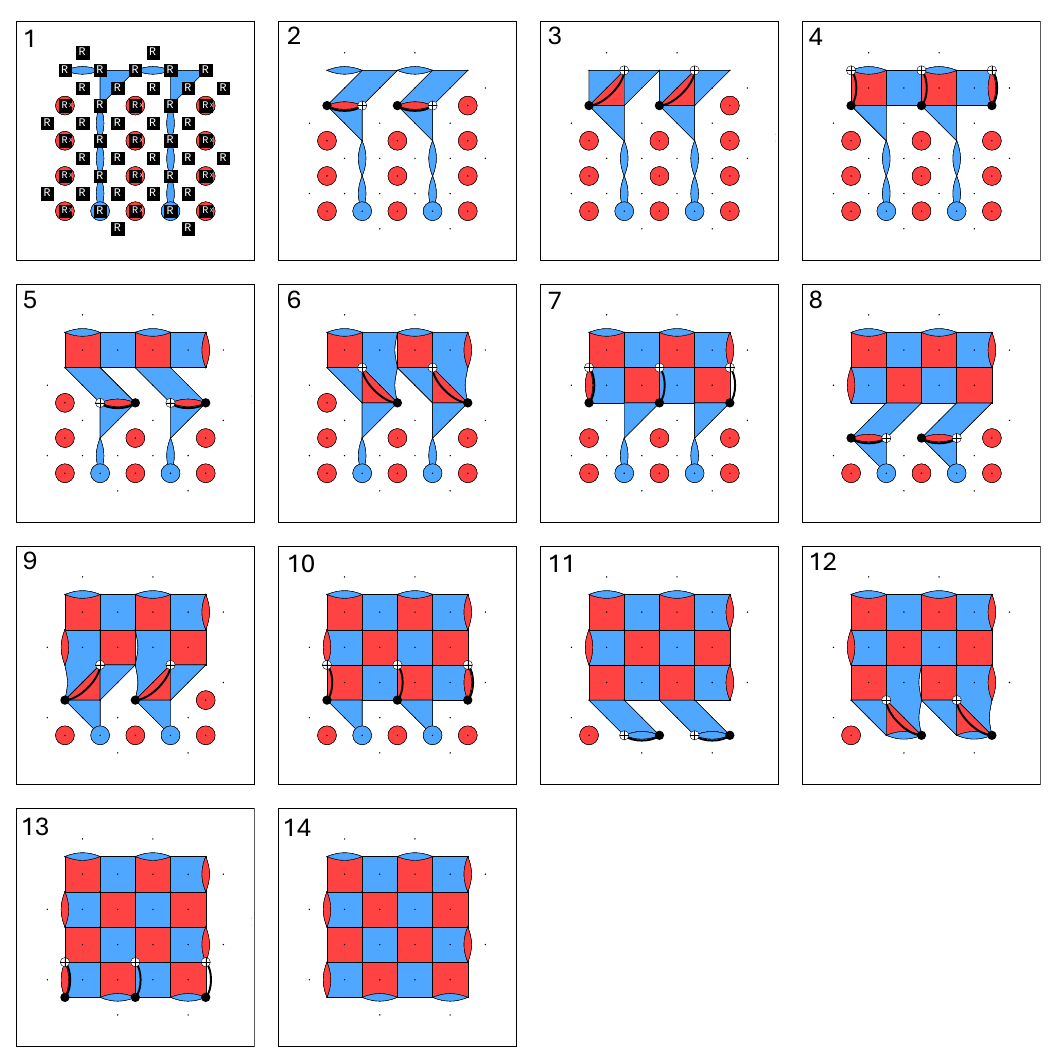}
    \caption{Unitary FT encoding circuit for the $d=5$ rotated surface code with full intra-plaquette connectivity. This encoding circuit assumes full connectivity among data qubits within each plaquette, eliminating the need for ancilla-mediated interactions. As a result, the total circuit depth is reduced to 13 time steps, compared to the 21 steps required in the ancilla-based version shown in Fig.~\ref{fig:circuit_ancilla}. This reduction is due to the simplification of the stabilizer-expanding circuits, which require two fewer gates per stabilizer, as illustrated in Fig.~\ref{fig:stab_circuits}.
}
    \label{fig:circuit_no_ancilla}
\end{figure*}

In this section, we generalize the unitary fault-tolerant (FT) scheme discovered for $d=3$ rotated surface code by a reinforcement learning agent in the previous work \cite{zen_quantum_2025} in the context of flag fault-tolerance.
The fundamental building blocks are the stabilizer-expanding circuits shown in Fig.~\ref{fig:stab_circuits} for the $X$-type stabilizers. The corresponding $Z$-type stabilizers are obtained by reversing the orientation of the CNOT gates and appropriately initializing the data qubits; see Appendix~\ref{appendix:circuits_plus} for details.
In practice, these circuits operate by selecting a \textit{pivot} qubit and transforming it into the desired stabilizer, following a construction analogous to the one used to prepare graph states 
\cite{liao_graph-state_2021,lang_minimal_2012,amaro_scalable_2020}, therefore full connectivity among all data qubits within a stabilizer plaquette is required. However, we propose an equivalent circuit in which the plaquette qubits are connected only to a common ancilla qubit, specifically the one assigned to measure the corresponding stabilizer during subsequent QEC cycles.
Importantly, the error propagation behavior in stabilizer-expanding circuits with or without ancillas is similar. In both cases, a single hook error can potentially compromise fault tolerance, as illustrated in Fig.~\ref{fig:stab_circuits}. This allows us to use either circuit architecture interchangeably, depending on the specific qubit connectivity available within each plaquette. 
Importantly, the circuits shown in Fig.~\ref{fig:stab_circuits} do not involve any measurements; the ancilla qubit used in the construction does not perform a measurement on the newly prepared state. Instead, the ancilla serves solely as a bridge between data qubits, as discussed in \cite{zen_quantum_2025}.

We now describe in detail the preparation of the logical state $|0\rangle_L$ in the rotated surface code. The encoding procedure for $|+\rangle_L$ is analogous and discussed in Appendix~\ref{appendix:circuits_plus}.
The explicit protocol for initializing $|0\rangle_L$ in the $d=5$ rotated surface code using plaquette ancillas is shown in Fig.~\ref{fig:circuit_ancilla}. First, all data and ancilla qubits are initialized in the state $|0\rangle^{\otimes n}$. Then, the $X$-type stabilizers are expanded row by row, with each stabilizer prepared using the circuits from Fig.~\ref{fig:stab_circuits}. For the first row, the choice of pivot qubits is arbitrary. However, in subsequent rows, the pivot qubits must differ from those used in the previous row.
A crucial constraint is that any CNOT gate applied to a qubit already involved in a previously prepared stabilizer must use that qubit only as a target, never as a control. The latter condition ensures that hook errors do not propagate into larger correlated errors, which could degrade the code’s fault distance. As a result, stabilizers are prepared serially along each of the $d-1$ rows (or columns), leading to a circuit depth proportional to $d$.
In Fig.~\ref{fig:circuit_no_ancilla}, we also show the encoding circuit under the assumption of full connectivity among data qubits within each plaquette. 
This version results in circuits with shorter depth due to the reduced number of gates required per stabilizer, i.e. three two-qubit gates instead of five for the four-qubit plaquettes. The same procedure applies to the unrotated surface code, as discussed in Appendix~\ref{appendix:circuits_plus}.

Some remarks regarding the circuits shown in Fig.~\ref{fig:circuit_ancilla} and Fig.~\ref{fig:circuit_no_ancilla} are in order. 
First, the gate schedule depicted is not unique; further parallelization across different rows is feasible. Namely, some gates in row $d+1$ can already be applied when some plaquettes in row $d$ are still pending. 
Nonetheless, the overall circuit depth remains proportional to $d$. The schedule presented here is chosen to make the row-by-row structure of the operations explicit and to clearly illustrate the precise gate sequencing on each plaquette.
Second, the encoding is \textit{distance-preserving}, meaning the fault distance matches the code distance when preparing $|0\rangle_L$ ($|+\rangle_L$) and correcting $X$ ($Z$) errors. However, when considering the complementary type of error—for example, $Z$ errors during the preparation of $|0\rangle_L$—the effective fault distance is bounded. The same applies to the measurement encoding as well due to the presence of measurement errors \cite{bluvstein_logical_2024}. Distance preserving and fault-tolerance of the encoding is further discussed in Appendix~\ref{appendix:ft_both}. Such a scenario arises, for instance, when preparing logical Bell pairs using transversal CNOTs \cite{bluvstein_logical_2024} or when initializing logical ancilla qubits for Steane or Knill type error correction protocols \cite{knill_scalable_2005,steane_active_1997,postler_demonstration_2024}.

\section{Numerical simulations}\label{sec:simulations}

\begin{figure*}
    \centering
     \includegraphics[width=1.0\linewidth]{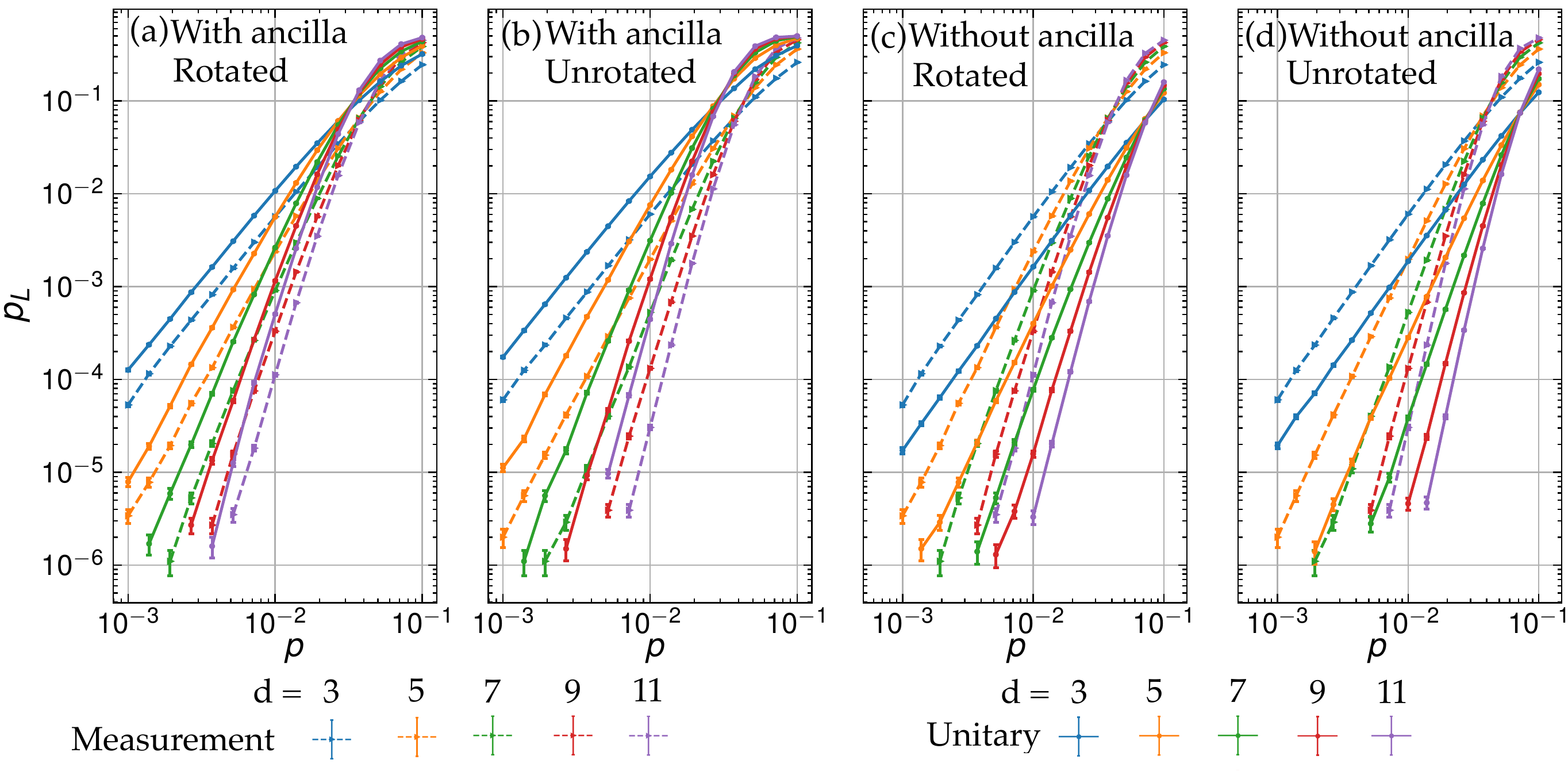}
    \caption{Logical error rate $p_L$ for preparing $|0\rangle_L$ for the unitary encoding schemes with, panels a) and b), and without, panels c) and d), ancilla for the rotated and unrotated surface code for $d=3,5,7,9,11$. $p$ is the physical error rate of gates and resets. Dashed lines are the measurement encoding (ME) while solid lines show different unitary encodings. For both types of codes, the encodings without ancillas perform better than the respective measurement encoding.}
    \label{fig:simulations}
\end{figure*}

\begin{table}
\begin{center}
\begin{tabular}{ |c|c|c|c| } 
 \hline
  & Gate count & Circuit depth & Measurements  \\ 
 \hline
 Rot. UEA & $(5t+3)(d-1)$ & $5(d-1)$ & No  \\ 
 \hline
 Rot. UE & $(3t+1)(d-1)$ & $3(d-1)$ & No  \\ 
 \hline
 Rot. ME & $2d(d-1)$ & $4$ & Yes  \\ 
 \hline
 UnRot. UEA & $(5d-2)(d-1)$ & $5(d-1)$ & No  \\ 
 \hline
 UnRot. UE & $(3d-2)(d-1)$ & $3(d-1)$ & No  \\ 
 \hline
 UnRot. ME & $2(2d-1)(d-1)$ & $4$ & Yes  \\ 
 \hline
\end{tabular}
\end{center}
\caption{Two-qubit gate count and circuit depth of the unitary encoding with ancilla (UEA) and without (UE) for the rotated (Rot.) and unrotated (UnRot.) surface code for odd code distance $d$. Here we use $t=\lfloor (d-1)/2\rfloor$. Gate count and circuit depth of measurement encoding (ME) is also shown.}
\label{tab:gate_count}
\end{table}

To evaluate the performance of our encoding scheme, we model imperfect gates using depolarizing noise channels applied immediately after each gate operation. 
Specifically, two-qubit gates are followed by a two-qubit depolarizing channel,
\begin{equation}
\mathcal{N}_2(\rho) = (1 - p)\rho + \frac{p}{15} \sum_{P} P \rho P,
\end{equation}
where \( P \in \{ \sigma \otimes \sigma \} \setminus \{I \otimes I\} \) and \( \sigma = \{I, X, Y, Z\} \) denotes the set of Pauli matrices. Preparation of the $|0\rangle$ and $|+\rangle$ single qubit states is performed via reset operations. Each preparation of the $|0\rangle$ state is followed by an error channel of the form: 
\begin{equation}
\mathcal{N}^Z_1(\rho) = (1 - p)\rho + p  X \rho X.
\end{equation}
Analogously, the preparation of the state $|+\rangle$ is followed by the error channel 
\begin{equation}
\mathcal{N}^X_1(\rho) = (1 - p)\rho + p  Z \rho Z.
\end{equation}

We benchmark our encoding scheme against the standard stabilizer measurement initialization scheme, which prepares the logical state \( |0\rangle_L \) (or \( |+\rangle_L \)) by first initializing all physical qubits in the product state \( |0\rangle^{\otimes n} \) (or \( |+\rangle^{\otimes n} \)), followed by a single round of \( X \)- (or \( Z \)-) stabilizer measurements. In the absence of noise, this procedure yields a state with all \( Z \)- (or \( X \)-) stabilizers equal to \( +1 \), while the conjugate stabilizers have a random value. Since error correction for \( |0\rangle_L \) (or \( |+\rangle_L \)) relies only on the measurement of \( Z \)- (or \( X \)-) stabilizers, a single round of fault-tolerant stabilizer measurements suffices to prepare a fault-tolerant logical state. 
However, when logical operations are subsequently applied, a full quantum error correction (QEC) cycle comprising \( d \) rounds of stabilizer measurements is required to maintain fault tolerance. In this work, we focus solely on the quality of the prepared Pauli eigenstates \( |0\rangle_L \) and \( |+\rangle_L \), and therefore limit our analysis to a single round of noisy stabilizer measurements. Given that measuring a weight-four (or weight-two) stabilizer requires four (or two) two-qubit gates, the total number of two-qubit gates needed for a single round is $2d(d-1)$ for the rotated surface code, see Table~\ref{tab:gate_count}.

We simulate three types of surface code encoding circuits: (i) measurement-based encoding (ME), (ii) unitary encoding with ancilla (UEA), and (iii) unitary encoding without ancilla (UE), for both the rotated and unrotated surface code. To evaluate the quality of the prepared logical states, each encoding circuit is followed by a perfect (noiseless) measurement of the data qubits and subsequent syndrome reconstruction and correction. This allows us to isolate the performance of the encoding circuit itself.
First we look at the gate count and circuit depth, see Table~\ref{tab:gate_count}. Every encoding scheme shown has a different number of entangling gates per stabilizer: 4 for ME, 5 for UEA, and 3 for UE (in the case of 4-qubit plaquettes). 
The circuit depth is the same for both realizations of the surface codes and unitary encodings, however the gate count of the rotated surface code is smaller due to the reduced number of stabilizers. Importantly, the gate count of the UE encoding is always smaller than the one of the ME encoding for $d>1$. 
Second, in Fig.~\ref{fig:simulations}, we compare the ME and UEA schemes for the preparation of the logical state $|0\rangle_L$. For both rotated and unrotated surface code, the ME scheme consistently yields a lower logical error rate than the UEA scheme, which can be attributed to its smaller number of gates. For example, for $d=5$ the ME and UEA encoding protocols for the rotated surface code need 40 and 52 gates, respectively. 
We also compare the ME and UE schemes. Here, the UE circuit outperforms ME for both the rotated and unrotated surface code, as expected given its lower gate count. 
For example, for $d=5$ the unrotated surface the ME and UE encoding schemes need 72 and 44 gates, respectively.
In some instances, we observe a reduction in logical error rate by up to an order of magnitude. Similar trends are observed for the preparation of the $|+\rangle_L$ state, as discussed in Appendix~\ref{appendix:circuits_plus}.
Since error propagation in each type of circuit is similar, specifically, hook errors are always kept under control, the differences in performance can be attributed mostly to the number of error locations in each circuit. 
Hence, avoiding the use of ancilla qubits considerably reduces the number of error locations, since ancilla qubits are themselves noisy and can introduce additional errors into the data qubits.

\section{Conclusions}\label{sec:conclusions}

We have presented a unitary encoding scheme for the surface code that is distance-preserving and requires only local connectivity. The circuit depth is $\mathcal{O}(d)$, as expected from fundamental bounds on entanglement generation in topological codes.
Our encoding scheme does not rely on measurements and classical feedback, making it particularly suitable for trapped-ion and neutral-atom platforms, where idling noise can be negligible and measurements are often considerably more noisy and slower than gates. 
The measurement-free encoding without ancillas outperforms the standard stabilizer-based encoding due to the reduced number of entangling gates per stabilizer and the absence of errors propagating from the ancilla qubits to the data qubits. 
In contrast, the unitary encoding with ancilla qubits performs worse. However, in a more realistic scenario that includes idling errors, replacing measurements with gates can reduce the overall protocol duration and the accumulation of idling errors, since measurements are usually considerably slower than gates.

Our work bridges the gap between unitary and non-unitary encodings of surface-code Pauli states. However, some important questions remain open. For instance, is it possible to extend the present encoding scheme to all-to-all connectivity and thereby further reduce the circuit depth? Since our encoding strategy relies on avoiding hook errors, it is not clear how to extend it in a distance-preserving manner in the presence of long-range gates. Nonetheless, this does not exclude the possibility that a different strategy for unitary encoding could be designed for all-to-all connectivity and achieve fault tolerance.  

Another open question is whether there exists a unitary encoding that simultaneously mitigates both types of hook errors. Such an encoding would be highly desirable for Steane-type error correction and for quantum algorithms supporting transversal logical gates.

\begin{acknowledgments}
We thank Josias Old for fruitful discussions. L.C. and M.M. gratefully acknowledge funding by the U.S. ARO Grant No. W911NF-21-1-0007. M.M. furthermore acknowledges funding from the European Union’s Horizon Europe research and innovation programme under grant agreement No 101114305 (“MILLENION-SGA1” EU Project), and the German Federal Ministry of Research, Technology and Space (BMFTR) as part of the Research Program Quantum Systems, research project 13N17317 (”SQale”), and MUNIQC-ATOMS (Grant No. 13N16070). This research is also part of the Munich Quantum Valley (K-4 and K-8), which is supported by the Bavarian state government with funds from the Hightech Agenda Bayern Plus. M.M. acknowledges funding from
the ERC Starting Grant QNets through Grant No. 804247, and from the European Union’s Horizon Europe research and innovation program under Grant Agreement No. 101046968 (BRISQ). 
L.C.~and M.M. also acknowledge support for the research that was sponsored by IARPA and the Army Research Office, under the Entangled Logical Qubits program, and was accomplished under Cooperative Agreement Number W911NF-23-2-0216. The views and conclusions contained in this document are those of the authors and should not be interpreted as representing the official policies, either expressed or implied, of IARPA, the Army Research Office, or the U.S. Government. The U.S. Government is authorized to reproduce and distribute reprints for Government purposes notwithstanding any copyright notation herein. M.M. acknowledges support from the Deutsche Forschungsgemeinschaft (DFG, German Research Foundation) under  Germany’s Excellence Strategy Cluster of Excellence Matter and Light for  Quantum Computing (ML4Q) EXC 2004/1 390534769. The
authors gratefully acknowledge the computing time provided to them at the NHR Center NHR4CES at RWTH
Aachen University (Project No. p0020074). This is
funded by the Federal Ministry of Education and Research and the state governments participating on the
basis of the resolutions of the GWK for national high
performance computing at universities.

We used the Python libraries NumPy \cite{harris2020array} and Stim \cite{gidney2021stim}. 
The Stim circuits for the simulations shown in Fig.~\ref{fig:simulations} and ~\ref{fig:simulations_plus} can be found in the Zenodo repository \href{https://doi.org/10.5281/zenodo.18014961}{https://doi.org/10.5281/zenodo.18014961}. 

\end{acknowledgments}

\bibliography{references,footnote}

\newpage
\clearpage
\onecolumngrid
\appendix

\section{Circuits for encoding $|+\rangle_L$}\label{appendix:circuits_plus}

\begin{figure}
    \centering
     \includegraphics[width=0.8\linewidth]{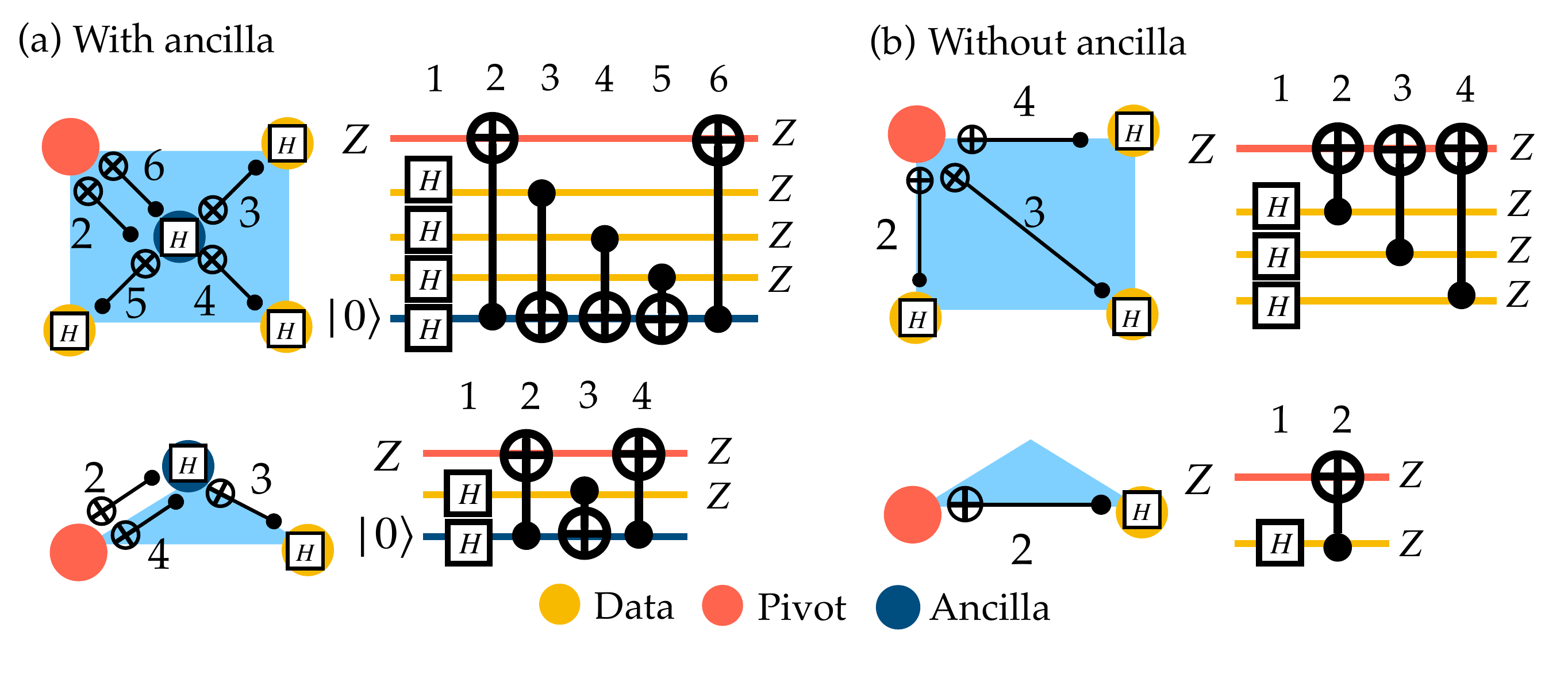}
    \caption{Circuits for expanding the $Z$ stabilizers. a) and b) Circuit for creating and $Z$ plaquette with and without ancilla in the middle, respectively. Each circuit assumes that the pivot qubit (positioned at the top of each diagram) is initialized in the state $|0\rangle$ while the others are initialized in $|+\rangle$ (shown as a Hadamard gate in this depiction). The circuit then transforms the pivot qubit into an $Z$-type stabilizer. Numbers indicate the order of the gates to avoid proliferation of dangerous hook errors. }
    \label{fig:circuit_ancilla_plus_picture}
\end{figure}

\begin{figure*}
    \centering
     \includegraphics[width=1.0\linewidth]{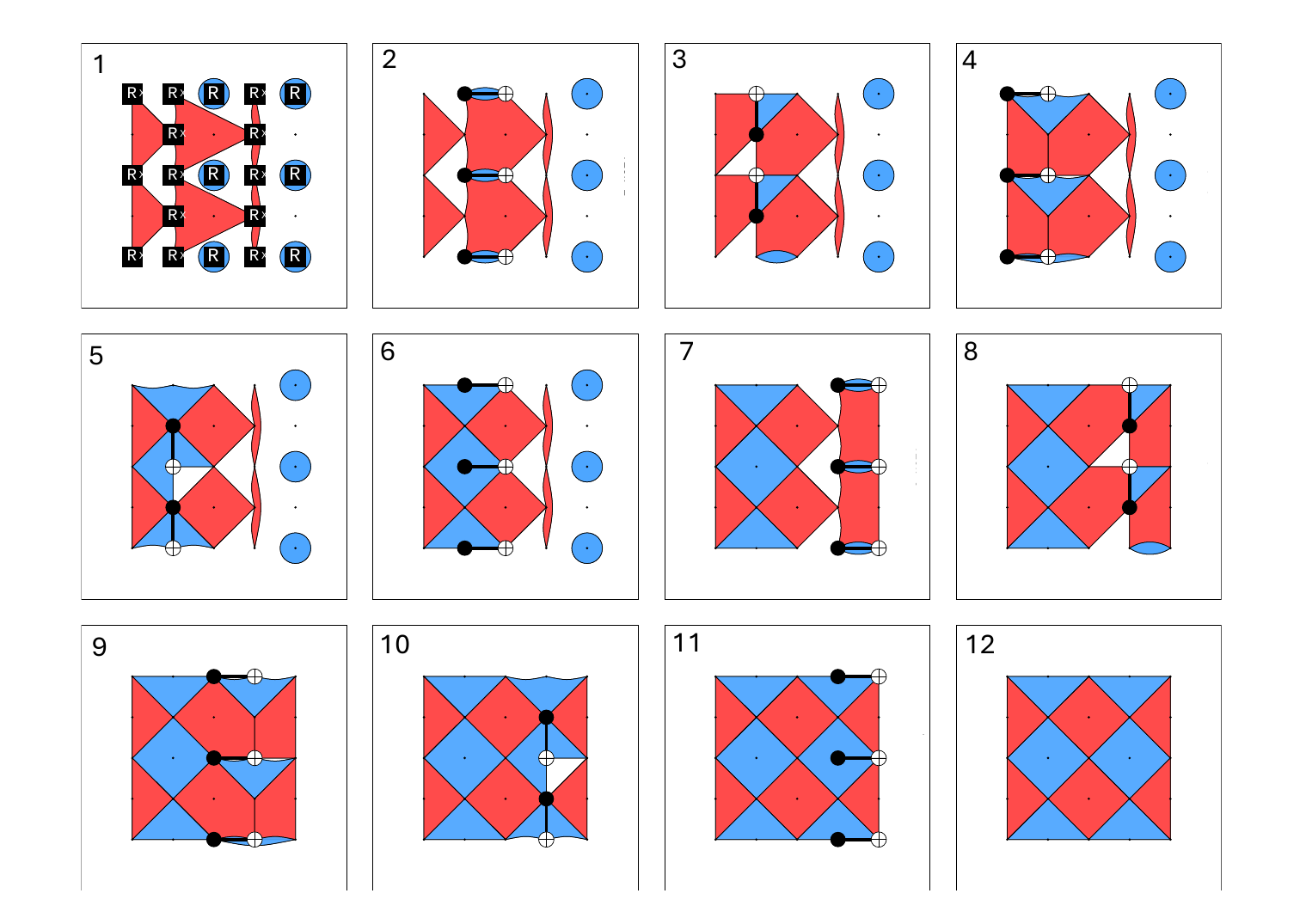}
    \caption{Unitary FT encoding for the $d=3$ unrotated surface code of $|+\rangle_L$. In this layout, the logical operators $X_L$ and $Z_L$ lie along the horizontal and vertical directions, respectively. Step 1 initializes the data and ancilla qubits in the states $|0\rangle$ or $|+\rangle$, allowing all pivot qubits to be selected and prepared in advance. Steps 2 through 6 prepare the $Z$-type stabilizers along the leftmost column. Each plaquette within the row is prepared in parallel. Steps 7 through 11 perform the preparation of $Z$ stabilizers on the second (rightmost) row. Step 12 depicts the completed state. At the end of the circuit, only correctable weight-two $Z$ errors can remain. }
    \label{fig:circuit_ancilla_plus}
\end{figure*}

\begin{figure*}
    \centering
     \includegraphics[width=1.0\linewidth]{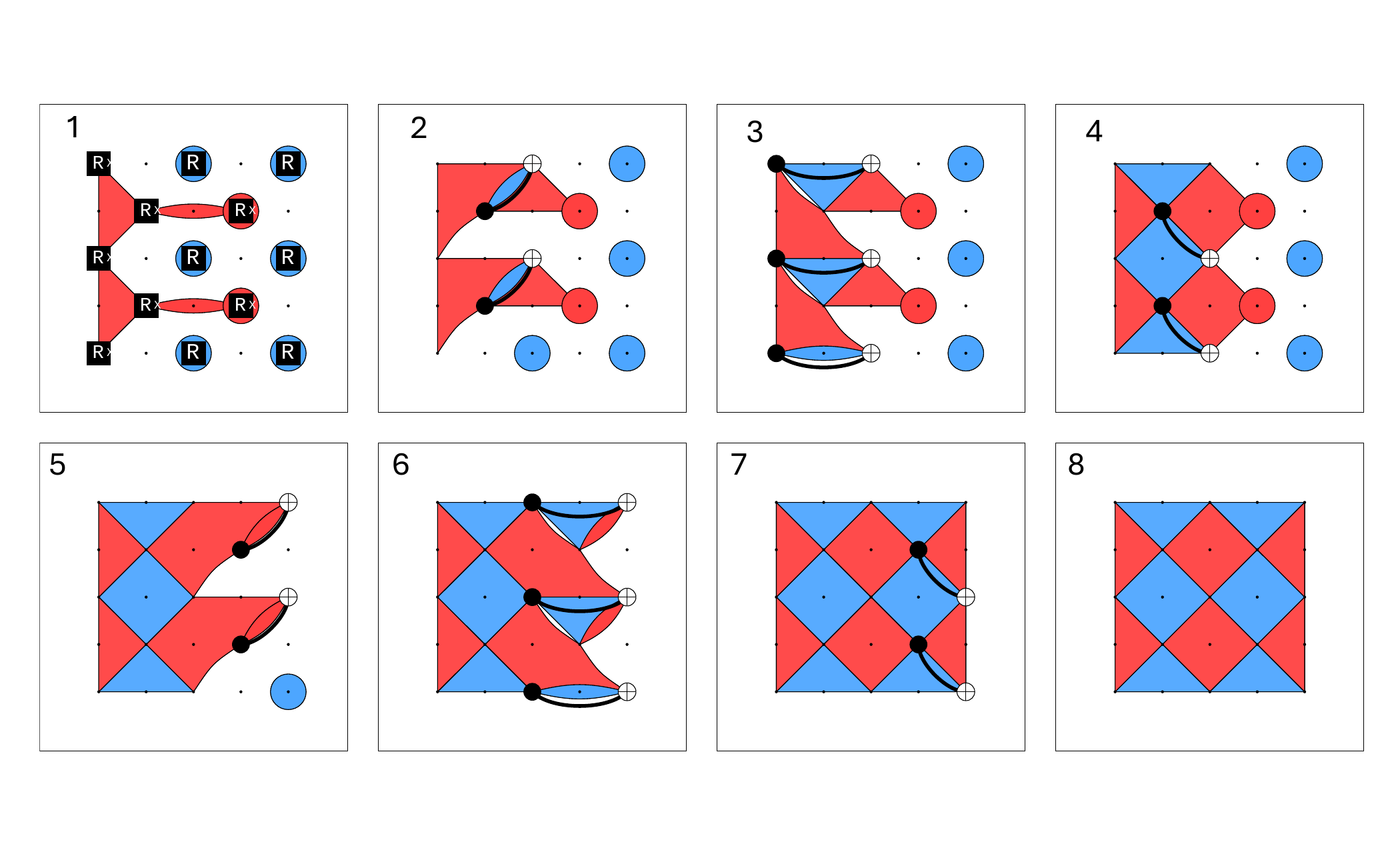}
    \caption{Unitary FT encoding circuit for the $d=3$ $|+\rangle_L$ in the unrotated surface code. This encoding circuit assumes full connectivity among data qubits within each plaquette, eliminating the need for ancilla-mediated interactions. As a result, the total circuit depth is reduced to 7 time steps, compared to the 11 steps required in the ancilla-based version shown in Fig.~\ref{fig:circuit_ancilla_plus}.}
    \label{fig:circuit_no_ancilla_plus}
\end{figure*}

\begin{figure}
    \centering
     \includegraphics[width=0.85\linewidth]{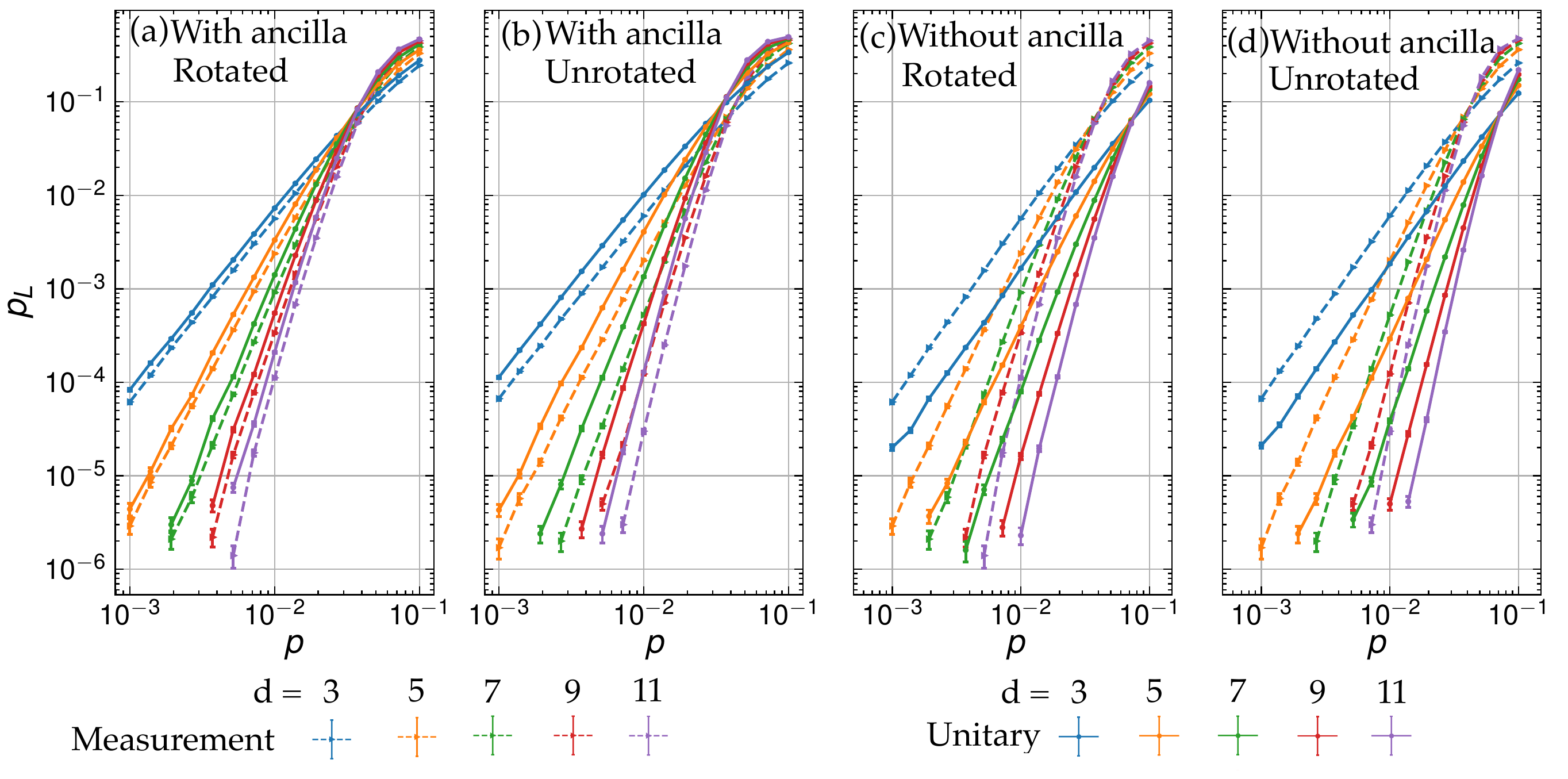}
    \caption{Logical error rate for preparing $|+\rangle_L$ for the different unitary encoding schemes with (a and b) and without (c and d) ancilla for the rotated and unrotated surface code for $d=3,5,7,9,11$. Dashed lines correspond to the measurement-based encoding while continuous lines show different unitary encodings. The qualitative behavior is similar as for the preparation of $|0\rangle_L$, see Fig.~\ref{fig:simulations}.}
    \label{fig:simulations_plus}
\end{figure}

In this section, we describe the unitary encoding scheme for preparing the logical state $|+\rangle_L$. It works analogously to the scheme explained in the main text for preparing $|0\rangle_L$. First, data qubits are initialized in $|+\rangle$, except for the pivot qubits, which are prepared in $|0\rangle$. Then, the $Z$ stabilizers are created using the circuits shown in Fig.~\ref{fig:circuit_ancilla_plus_picture}. Stabilizers within each column are prepared in parallel, while the columns themselves are prepared sequentially. The complete circuits for the $d=3$ unrotated surface code are presented in Fig.~\ref{fig:circuit_ancilla_plus} (with ancillas) and Fig.~\ref{fig:circuit_no_ancilla_plus} (without ancillas).  
Encoding $|0\rangle_L$ and $|+\rangle_L$ in the unrotated surface code proceeds in the same way as in the rotated version. The only difference is the presence of weight-3 stabilizers at the boundaries instead of weight-2, which may introduce hook errors. However, the unrotated surface code has the property that only combinations of two-qubit errors within each plaquette can potentially induce a logical error~\cite{manes_distance-preserving_2025,dennis_topological_2002,old_fault-tolerant_2025,jandura_surface_2024}. The reason is that each plaquette contains only two qubits along the direction of the logical operators, so only one specific hook error must be avoided. As shown in Fig.~\ref{fig:stab_circuits}, there is only one possible harmful hook error; thus, the task reduces to ordering the gates such that this error does not affect qubits along the corresponding logical operator. This strategy is effective for every code that, like the surface code, tolerates certain weight-2 errors.  

Finally, we present simulations for the preparation of the $|+\rangle_L$ state in both the rotated and unrotated surface codes. Figure~\ref{fig:simulations_plus} shows the results for the encoding schemes with and without ancillas. As discussed in the main text for $|0\rangle_L$, the unitary encoding scheme without ancillas achieves a substantially lower logical error rate than the measurement-based encoding. The scheme with ancillas has a higher logical error rate than the measurement-based encoding, but the difference is not as pronounced as in the case of $|0\rangle_L$.

\section{Fault-tolerance against both types of errors}\label{appendix:ft_both}

\begin{figure}
    \centering
     \includegraphics[width=0.7\linewidth]{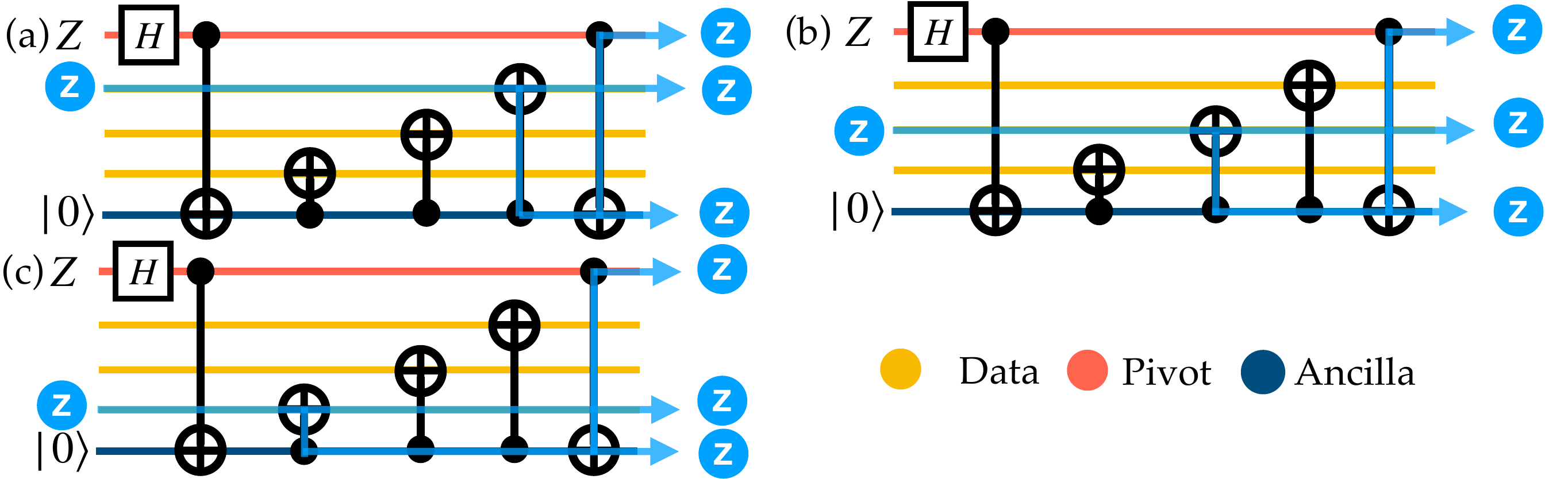}
    \caption{Hook $Z$ errors in the expanding $X$ stabilizer circuits. A single $Z$ error on the data qubits (excluding the pivot) can lead to a different hook error each. Besides, it will not flip state of the ancilla, making it undetectable even when using ancillas as flags.}
    \label{fig:hook_z_errors}
\end{figure}

As discussed in the main text, the unitary encoding scheme we propose is able to preserve the code distance with respect to one type of error at a time. For example, when preparing $|0\rangle_L$, some high-weight $Z$ errors may still occur. In this section, we explain why such errors cannot be fully avoided. As shown in Fig.~\ref{fig:hook_z_errors}, the circuits for preparing the $X$ stabilizers can generate hook $Z$ errors at three different non-equivalent locations. In fact, any weight-2 error involving the pivot qubit may arise. This imposes strong limitations on the fault distance for $Z$ errors, since such errors can propagate through the entire code via the entangling gates applied in later stages. An analogous situation occurs for $X$ errors when preparing $|+\rangle_L$.

\end{document}